\documentclass[preprint,12pt]{elsarticle}

\usepackage{amssymb}

\journal{$^*$Corresponding author. Email: yangli@iie.ac.cn}

\begin{document}

\begin{frontmatter}



\title{Unicity Distance of Quantum Encryption Protocols}

\author{Chong Xiang}
\author{Li Yang$^*$}

\address{State Key Laboratory of Information Security, Institute of Information Engineering, Chinese Academy of Sciences, Beijing 10093, China}

\begin{abstract}
Shannon presented the concept `unicity distance' for describing the security of secret key encryption protocols against various ciphertext-only attacks. We develop this important concept of cryptanalysis into the quantum context, and find that there exist quantum encryption protocols reusing a short key have infinite unicity distance.
\end{abstract}




\end{frontmatter}


\newtheorem{theorem}{Theorem}
\newtheorem{lemma}[theorem]{Lemma}
\newtheorem{conjecture}[theorem]{Conjecture}
\newtheorem{corollary}[theorem]{Corollary}
\newtheorem{definition}{Definition}
\newtheorem{proposition}[theorem]{Proposition}
\newtheorem{property}{Property}

\section{Introduction}
Shannon provided the concept of unicity distance in 1949\cite{sha49} . He showed a way of calculating approximately how many ciphertexts were necessary to recovery the unique key for the classical encryption protocols, the length of ciphertexts needed was determined by the entropy of the key space and the redundance of the plaintext space.
This parameter can be used to measure the number of times one key can be reused securely, while the length of ciphertexts encrypted by same key is not more than the unicity distance, this key can not be confirmed under the Known Ciphertext Attack(KCA). According to Deavours, the unicity distance of DES is worked out as 8.2 ASCII characters or 66 bits\cite{Dea77,Sch96} .  Mao and Wu\cite{Mao07} used the unicity distance to analyse the security of hashing algorithms, Liu, Zhang and Li\cite{Liu10} then develop it into multimedia hash scheme.

The quantum key distribution has been proved to be unconditionally secure and applied in practice\cite{Eke91,Lo95,Deu96,May96_QKD,May01}. Though it is seem that quantum one-time pad(which usually called private quantum channel\cite{Amb00,Amb04,Boy03}) can be achieved theoretically with the quantum key distribution, we still need to consider the repetition of key for the well-known low efficiency of the quantum key distribution. The unicity distance of quantum encryption protocols represents the theoretical limitation of how many times a key can be reused in quantum context. The unicity distance of a quantum encryption protocol can be used to measure its efficiency and the security in practice.

\section{Preliminaries}
\subsection{Private key encryption protocol}
The private key encryption protocols can be divided into five different kinds based on that if the plaintext space, the key space, and the algorithm are belong to classical or quantum context, we present the classification as follows:

\begin{enumerate}
  \item The plaintext and key are both classical, the algorithm is also classical.
  \item The plaintext and key are both classical, but the algorithm is a quantum one(CCQ).
  \item The plaintext is classical, but the key and the algorithm are quantum ones.
  \item The plaintext is quantum, the key is classical, but the algorithm is a quantum one.
  \item The plaintext, the key, and the algorithm are quantum.
\end{enumerate}

Within this classification, the first kind of protocol is the protocol generally called classical encryption protocol, and the other four kinds of protocol are all called quantum encryption protocols. This classification is not based on the ciphertext space, because we think the ciphertext space's should be coincident with the algorithm. Furthermore, we think the classical algorithm can not be played on quantum states, so the algorithm of a protocol with quantum states must be quantum operation. More detailed discussions about this classification will be presented in our further work.

Shannon's unicity distance is presented in classical context. In this paper, we discussed about the unicity distance of the second kind of protocol with classical plaintext and key space and quantum algorithm, we call it unicity distance of CCQ encryption protocol.

\subsection{Information-theoretic security}
In classical cryptography, the information-theoretic security is suggested by O. Goldrich \cite{Gold04} as follows:

\begin{definition} An encryption is information-theoretically secure if for every circuit family \{$C_n$\}, every positive polynomial $p(\cdot)$, all sufficiently large $n$'s, and every $x, y$ in plaintext space:
\begin{eqnarray}
\Big|\textrm{Pr}[C_n(G(1^n), E_{G(1^n)}(x))=1]-\textrm{Pr}[C_n(G(1^n), E_{G(1^n)}(y))=1]\Big|<\frac{1}{p(n)},
\end{eqnarray}
where $G$ is a key generation algorithm.
\end{definition}

We have suggested a definition of information-theoretic security of quantum encryption\cite{Cho12} as follows:

\begin{definition}A quantum encryption is information-theoretically secure if for every quantum circuit family \{$C_n$\}, every positive polynomial $p(\cdot)$, all sufficiently large $n$, and every $x, y$ in plaintext space:
\begin{eqnarray}\label{ITS}
\Big|\textrm{Pr}[C_n(G(1^n), E_{G(1^n)}(x)=1]-\textrm{Pr}[C_n(G(1^n), E_{G(1^n)}(y))=1]\Big|<\frac{1}{p(n)},
\end{eqnarray}
where the encryption algorithm $E$ is a quantum algorithm, and the ciphertext $E(x), E(y)$ are quantum states.
\end{definition}

And the following theorem shows a way to prove a quantum encryption scheme is information theoretically indistinguishable:
\begin{theorem}\label{information}
For every plaintexts $x$ and $y$, let the density operators of cipher states $G(1^n)(x)$ and $G(1^n)(y)$ are $\rho_x$ and $\rho_y$, respectively. A quantum private-key encryption is said to be information theoretically indistinguishable if for every positive polynomial $p(\cdot)$ and every sufficiently large $n$'s,
\begin{eqnarray}
D(\rho_x,\rho_y)<\frac{1}{p(n)}.
\end{eqnarray}
\end{theorem}

\subsection{Unicity distance}
The Shannon's unicity distance is a crucial quantify for evaluating the security of a private key encryption protocol. There are some paraments related to it:
\begin{enumerate}
  \item $n_1(y)$: while the length of a given ciphertext $y$ reachs $n_1(y)$, the attacker can determine the key from it.
  \item $n_2$, $n_2=\sum_yp(y)n_1(y)$: it is the average length of ciphertext necessary for the attacker to determine the key.
  \item $n_3$(a lower bound of $n_1$): if the attacker wants to obtain the key, the length of ciphertext required must not less than $n_3$.
  \item $n_4$(an upper bound of $n_1$): when the attacker owns a ciphertext with length $n_4$, he can determine the key.
  \item $n_5$(a lower bound of $n_2$): according to information theory and properties of function, $n_2$ is proved bigger than $n_5$.
\end{enumerate}

It must be noticed that the generally proved unicity distance $\frac{H(K)}{R_L\log_2|P|}$ is indeed a $n_5$, so people always need more ciphertexts to get the unique encryption key.

Based on the definition of unicity distance, it can be seen that if the ciphertext eavesdropper Eve owned is longer than the unicity distance, it can only be decrypted into a meaningful plaintext with the unique encryption key, but decrypt into meaningless message with any other keys. Which means the ciphertext sets encrypted from the same set of meaningful plaintexts with different keys must be almost mutually disjoint sets.

Classical encryption protocols have a property that its decryption process is certain, which means every time we decrypt the same ciphertext with the same key will result the same plaintext; however, the quantum encryption protocols have a certain decryption result with the unique right key but a uncertain decryption result with a random bit-string. It means that a quantum encryption protocol may not have unicity distance even if its ciphertext sets encrypted from the same set of meaningful plaintext with different keys are almost mutually disjoint sets.

Analogy with the proof for classical unicity distance, we can give a similar proof for quantum unicity distance:

\begin{lemma}\label{lem1}
Let the quintuple $(P, C, K, E, D)$ is a CCQ encryption scheme, then we have $$S(\rho_K|\rho_C)=S(\rho_K)+S(\rho_P)-S(\rho_C),$$
where $\rho_P, \rho_K, \rho_C$ are the density operators of the plaintext space, key space, and ciphertext space respectively, $S(\rho)$ expresses the Von Neumann entropy.
\end{lemma}
{\bf Remark:} {\sl The elements of a classical space can be regarded as quantum states coded by a set of standard orthogonal bases, so we can get the density operator of a classical space.}

{\bf Proof:} According to the property of Von Neumann entropy, we have
$$S(\rho_K|\rho_C)=S(\rho_{K,C})-S(\rho_C).$$

In order to decrypting the ciphertext successfully, the encryption algorithm is always a unitary operator, which is named as $U_E$. So
$$U_E(\rho_P\otimes\rho_K)U_E^\dag=\rho_{K,C},$$

then
$$S(\rho_K|\rho_C)=S(\rho_{K,C})-S(\rho_C)=S(U_D(\rho_P\otimes\rho_K)U_D^\dag)-S(\rho_C)$$
$$=S(\rho_P\otimes\rho_K)-S(\rho_C)=S(\rho_K)+S(\rho_P)-S(\rho_C).$$
Thus the lemma follows.  $\Box$

We define $P=\{0,1\}^l$,$C$ is a $l$-qubits space, let the redundancy of plaintext be $R_L$, so we get:
\begin{theorem}\label{the1}
  Let the quintuple $(P, C, K, E, D)$ is a CCQ encryption scheme, given a long enough (length is $n$) ciphertext, the expectation $\overline{S}_n$ of spurious key satisfies
  \begin{eqnarray}
    S(\rho_K|\rho_{C^n})\leq\log(\overline{S}_n+1).
  \end{eqnarray}
\end{theorem}
{\bf Proof:} For a given $Y\in C^n$, the number of spurious key is $|K(Y)|-1$, so
$$\overline{S}_n=\Sigma_{Y\in C^n}p(Y)(|K(Y)|-1)=\Sigma_{Y\in C^n}p(Y)|K(Y)|-1$$
Based on Lemma \ref{lem1}
$$S(\rho_K|\rho_{C^n})=S(\rho_K)+S(\rho_{P^n})-S(\rho_{C^n}),$$
here $S(\rho_{C^n})\leq \log(2^{l\times n})=l\times n$, $S(\rho_K)=H(K)$, and while plaintext is long enough,
$S(\rho_{P^n})=H(P^n)=n(1-R_L)\times l$.

Then we get
\begin{eqnarray}\label{left}
  S(\rho_K|\rho_{C^n})\geq H(k)-nlR_L.
\end{eqnarray}

On the other side, for every given $Y\in C^n$, $S(\rho_K|\rho_Y)\leq log|K(Y)|$, so
\begin{eqnarray}\label{right}
  S(\rho_K|\rho_{C^n})&\leq& \sum_{Y\in C^n}p(Y)log|K(Y)|\nonumber\\
   &\leq& log\sum_{Y\in C^n}p(Y)|K(Y)|=log(\overline{S}_n+1).
\end{eqnarray}
Thus the theorem follows.  $\Box$

Then we can get $$\overline{S}_n\geq\frac{2^{H(K)}}{2^{nlR_L}}-1$$ as the result of theorem \ref{the1}, which means if $\overline{S}_n=0$, we can get \begin{eqnarray}\label{ineqn2}
  n\geq\frac{H(K)}{lR_L}.
\end{eqnarray}
However, this result is not relevant to the quantum ciphertext-only attack for the properties of quantum information is rather different from classical information in very basic aspect. Obtaining a classical message is the same as knowing what it is, but it does not mean right in a quantum situation. Once we get a unknown quantum state which would be one of some non-orthogonal states, the properties of quantum mechanical determine that we can not accurately distinguish it. Here we must distinguish two relations between quantum states and people:
\begin{enumerate}
  \item ``A quantum state in one's mind":

A quantum state in one's mind indicates one knows the mathematic expression of the state, or the process preparing it exactly.

  \item ``A quantum state in one's hand":

A quantum state in one's hand indicates one owns the quantum register which contains the qubits. Generally speaking, one cannot know what the state is according to basic principles of quantum theory.
\end{enumerate}

In practice, the attacker can only get a quantum register with the ciphertext state, rather than a mathematical expression of the state, so we need to consider the concept of quantum unicity distance(QUD) for CCQ encryption protocols under this situation, and it must be bigger than Ineq. (\ref{ineqn2}).

\section{Quantum extension of unicity distance for CCQ encryption protocol(QUD)}
According to a CCQ encryption scheme, assume the key is $k_0$ with length $n$, the plaintext is $i\in P$, then the encryption process $\mathcal{E}_{k_0}$ is shown as:
\begin{eqnarray}
  \mathcal{E}_{k_0}(|i\rangle\langle i|)
  &=&Tr_k\left(U_E|i\rangle\langle i|\otimes|k_0\rangle\langle k_0|U_E^{\dag}\right)\nonumber\\
  &=&\sum_{k\in K}\langle k|U_E|i\rangle\langle i|\otimes|k_0\rangle\langle k_0|U_E^{\dag}|k\rangle.
\end{eqnarray}

So $E_{kk_0}=\langle k|U_E|k_0\rangle$ is the operation element for $\mathcal{E}_{k_0}$, it satisfies that:
$$\sum_{k\in K}E_{kk_0}^\dag E_{kk_0}=I, \sum_{k_0\in K}E_{kk_0} E_{kk_0}^\dag=I.$$
Where the first equality is the completeness condition, and the second equality is a new relation for the encryption process.

We rewrite the expression as:
\begin{eqnarray}
  \mathcal{E}_{k_0}(|i\rangle\langle i|)=\sum_{k\in K}E_{kk_0}|i\rangle\langle i|E_{kk_0}^\dag.
\end{eqnarray}

The density operator of ciphertext for sender then is shown as:
\begin{eqnarray}
  \rho_A=\mathcal{E}_{k_0}(|i\rangle\langle i|)=\sum_{k\in K}E_{kk_0}|i\rangle\langle i|E_{kk_0}^\dag,
\end{eqnarray}
the density operator of ciphertext for receiver is shown as:
\begin{eqnarray}
  \rho_B=\mathcal{E}_{k_0}(\sum_{i\in P}p_i|i\rangle\langle i|)=\sum_{k\in K}\sum_{i\in P}p_iE_{kk_0}|i\rangle\langle i|E_{kk_0}^\dag,
\end{eqnarray}
and the density operator of ciphertext for attacker is shown as:
\begin{eqnarray}
  \rho_E=\sum_{k_0\in K}q_{k_0}\mathcal{E}_{k_0}(\sum_{i\in P}|i\rangle\langle i|)=\sum_{k_0\in K}\sum_{k\in K}\sum_{k\in K}p_iq_{k_0}E_{kk_0}|i\rangle\langle i|E_{kk_0}^\dag.
\end{eqnarray}

\subsection{\rm QUD1}\label{sec1}
The accessible information for the attacker with the quantum ciphertext in hand is bounded by Holevo $\chi$ quantity of the state:
\begin{eqnarray}
  H(M : I)\leq \chi,
\end{eqnarray}
where $H(M : I)$ is the information achieve from the ciphertext. So we can use $\chi$ to analyze the security of the quantum encryption protocol. While the length of ciphertexts is $n$, we can get
\begin{eqnarray}
  \chi(n)=S(\rho_E)-\sum_{i\in P}q_{i}S(\rho_{i}),
\end{eqnarray}
where $S(\cdot)$ is the Von Neumann entropy, $\rho_E$ is the density operator of ciphertext for attacker, and
\begin{eqnarray}
  \rho_{i}=\sum_{k_0\in K}\sum_{k\in K}p_{k_0}E_{kk_0}|i\rangle\langle i|E_{kk_0}^\dag.
\end{eqnarray}
where probabilities $q_i$ is determined by the plaintext space, and $p_{k_0}$ is determined by the key space.

Based on the theorem about the upper bound on the entropy of a mixture of quantum states \cite{Nie03}, We can have
\begin{eqnarray}
  \chi(n)=S(\rho_E)-\sum_{i\in P}q_{i}S(\rho_{i})\leq H(q_i)=n(1-R_L),
\end{eqnarray}
then the information of the key is limited by
\begin{eqnarray}
  S(\rho_E)-\chi(n)=\sum_{i\in P}q_{i}S(\rho_{i}).
\end{eqnarray}
Similarly, since $\sum_{k\in K}E_{kk_0}|i\rangle\langle i|E_{kk_0}^\dag$ is a pure state(while a particular plaintext and a key is given, the ciphertext is a pure state), we have
\begin{eqnarray}
  S(\rho_i)\leq H(p_{k_j})+\sum_{k_0\in K}p_{k_0}S\left(\sum_{k\in K}E_{kk_0}|i\rangle\langle i|E_{kk_0}^\dag\right )=H(p_{k_j})=H(K),
\end{eqnarray}
then
\begin{eqnarray}
  S(\rho_E)-\chi(n)\leq \sum_{i\in P}q_{i}H(K)=H(K).
\end{eqnarray}

It can be seen that, while reusing private key $k_0$ to encrypt plaintexts, suppose the length of the ciphertexts is $n$, $S(\rho_E)-\chi(n)$ will be increasing as the length of plaintext encrypted cumulated, though $H(K)$ is keeped the same. So there must be a bound $n_0$ which satisfies that the accessible information of $K$ is almost $H(K)$ in a range of errors less than any given small amounts. With this knowledge we give the first definition of QUD as follows:

\begin{definition}[QUD1]For a quantum encryption protocol, we call $n_0$ is the quantum unicity distance if it is the lower bound of the length of ciphertext, which satisfies:
\begin{eqnarray}
  S(\rho_E)-\chi(n_0)\geq H(K)-\epsilon,
\end{eqnarray}
\end{definition}

This definition means that if the accessible information of the key from the quantum ciphertext is near sufficiently the entropy of the key, the key can be determined. Here we should point out that, the classical unicity distance can also be explained in this way. From Ineq. (\ref{ineqn2}), we can get $nl-nl(1-R_L)\geq H(K)$, which is also means the accessible information of the ciphertext is greater than the information of plaintext and key, as the accessible information of a classical bit string is equal to its number of bits. However, there is an important difference between QUD1 and the definition of classical unicity distance: the classical unicity distance is always limited if the redundancy of plaintext is not zero, while the QUD1 may be unlimited for the same plaintext.

\subsection{{\rm QUD2}}
We can define the unicity distance of CCQ encryption protocol referring to Shannon's unicity distance, it turn out this result: the spurious key of a quantum ciphertext $Y$ may be defined as the bit-string with which decrypting $Y$ result a meaningful plaintext. But under quantum mechanical, the result of the decryption with a random bit-string should be have many possibilities, then the spurious key must make every possibility of the result being a meaningful plaintext, which is difficult to satisfy; on the other hand if we decrypt ciphertext with a random bit-string, the most possible condition should have some meaningful results and others meaningless. So we need to give a new definition for unicity distance of CCQ encryption protocol.

\begin{definition}The spurious key of the unicity distance of CCQ encryption protocol is the bit-string except the unique key $k_0$ with which decrypting ciphertext $Y$ may result a meaningful plaintext with probability more than $1-\delta$. Let's define set $Z$ as the set of meaningful plaintexts, $M$ is the measurement, the spurious key of $Y$ can be represented as follow:
    \begin{eqnarray}K(Y)=\{k\in K, k\neq k_0|{\rm Pr}(M(D_k(Y))\in Z)>1-\delta\}.\end{eqnarray}
\end{definition}

\begin{definition}[QUD2]For a quantum encryption protocol, if there exists a $N$ satisfying that while the length of ciphertext is bigger than $N$, the amount of quantum spurious key is 0, we will call $N$ the unicity distance of CCQ encryption protocol.
\end{definition}

Notice that a classical ciphertext can be used any times to verify all keys but a quantum ciphertext can only be used once, so unicity distance of CCQ encryption protocol should contain two components: the length of ciphertext with which we can verify if a bit-string is the key and the size of the key space. These will be addressed in more detail later in this article.

\subsection{{\rm QUD3}}
This definition of QUD2 is suited for deriving the unicity distance of a given quantum encryption protocol, but it is difficult to derive the unicity distance of a formalized quantum encryption protocol. Therefore we give another definition of unicity distance of CCQ encryption protocol.

Let $n$ is the length of quantum ciphertext $Y_n$, the amount of its quantum spurious key is:
    \begin{eqnarray}|K(Y_n)|=\sum_k\sum_{|\phi_i\rangle\in Z_n}\langle\phi_i|D_k(Y_n)|\phi_i\rangle=\sum_kTr({\rm P}(n)D_k(Y_n)),\end{eqnarray}
where $D_k$ is decryption transformation with key $k$, $Z_n$ contains all the elements with length $n$ of $Z$, and ${\rm P}(n)$ is the projection operator of $Z_n$. Then the unicity distance is defined as follow:

\begin{definition}[QUD3]For a quantum encryption protocol, if there exists a $N$ satisfying that while the length of ciphertext $n$ is bigger than $N$, $|K(Y_n)|=1$, we will call $N$ the unicity distance of CCQ encryption protocol.
\end{definition}

\section{Examples and discussions}
\subsection{Quantum encryption protocols with finite unicity distance(QUD2)}
Probabilistic encryption\cite{God84} is also called randomized encryption. A process of probabilistic encryption $E$ can be shown as follow:
\begin{eqnarray}E: V_m\times K\times R\rightarrow V_n,\end{eqnarray}
where $V_m$ and $V_n$ are plaintext space and ciphertext space with $m\leq n$, $K$ is the key space and $R$ is a set of random numbers.

The probabilistic encryption protocols maps one plaintext into different ciphertexts with the same key, so it can resist chosen plaintext attack better, and increase the effective size of the plaintext space.

Here two quantum encryption protocols are both probabilistic protocols. We will give their unicity distance based on QUD2.

\subsubsection{The protocol A}
For simplicity, we consider bitwise encryption. We use key $k\in\{0,1\}^l$ to encrypt a bit $x$ under following protocol:

The encryption: we randomly choose a bit-string $b=b_1,b_2,\cdots,b_l$, satisfies $b_1\oplus\cdots\oplus b_l=x$. The bit $x$ is
encrypted to be $|\psi_1\rangle\cdots|\phi_l\rangle$. Here
\begin{eqnarray}|\phi_i\rangle=\left\{\begin{array}{ll}
                                                    |0\rangle  &(k_i=0,b_i=0)\\
                                                    |1\rangle  &(k_i=0,b_i=1)\\
                                                    |+\rangle  &(k_i=1,b_i=0)\\
                                                    |-\rangle  &(k_i=1,b_i=1)
                                                    \end{array}\right.\end{eqnarray}

The decryption: we measure each qubit with bases determined by $k_i$, which will result $b_i$, then $b=b_1\oplus\cdots\oplus b_l$.

The security of this protocol is proved in \cite{Yan12}.

For the convenience of analysis, we assume the plaintext is the codewords being encoded with error correcting code(ECC) $C$, here $C$ is a $(m,k)$-ECC which can correct $t$-bit error. Then the minimum distance between each $m$-bit codewords is $2t+1$. Let the length of key is $m\times l$, make every $l-$bit used to encrypt one bit of plaintext as a group, so the key can be divided into $m$ groups. For any $m\times l$-bit-string, we can divide it into $m$ groups similarly. It is easy to find that once there exists any differences between the string and the key in one group, the decryption of this group will result of 0 or 1 with both probability of $50\%$, so we may just consider the number of groups with differences, named $m_0$. Let $p_e$ is the probability of checking one $m\times l$-bit-string.

\begin{enumerate}
    \item
     While $m_0<2t+1$, one $m\times l$-bit-string can not be checked out if and only if every group of string with error results out the right plaintext with probability $50\%$. When checking this kind of key with ciphertext whose length is $n\times m$, \begin{eqnarray}p_e=1-(\frac{1}{2})^{m_0\times n}.\end{eqnarray}

    \item
    While $m_0\geq2t+1$, an error key may result out $2^{m_0}$ different ciphertexts, there are at least $2^{2t}$ invalid ciphertexts. When checking this kind of key,
    \begin{eqnarray}p_e>1-(1-2^{2t-m_0})^n.\end{eqnarray}
\end{enumerate}

Let security parameter is $p_0=1-\delta$, if we want to satisfy $p_e\geq p_0$ with every $m_0$: if $m_0<2t+1$, it turns out $n=-\frac{\ln(1-p_0)}{m_0\ln2}$; else it turns out $n=\frac{\ln(1-p_0)}{\ln(1-2^{2t-m_0})}$.

Note that we need to check every key with the maximum $n$, so
\begin{eqnarray}n_0=\max_{m_0}{\{-\frac{\ln(1-p_0)}{m_0\ln2},\frac{\ln(1-p_0)}{\ln(1-2^{2t-m_0})}\}}=\frac{\ln(1-p_0)}{\ln(1-2^{2t-m})}.\end{eqnarray}

Then the unicity distance for $p_0$ is $N=n_0\times 2^{m\times l}$.

It was proved that $m\times l$-bits key just can encrypt $\frac{1}{2}m\times l$-bits plaintext\cite{Amb00,Boy03}. The gap between $\frac{1}{2}m\times l$ and $N$ is much bigger than that of classical encryption protocols.

\subsubsection{The protocol B}
We first consider bitwise encryption. Using the following scheme to encrypt every bit with key $k\in\{0,1\}^l$.

{\bf Encryption:} Alice chooses $i\in\{0,1\}^l$ randomly. A bit $x$ is encrypted to be
\begin{eqnarray}
\frac{|i\rangle+(-1)^{x}|i\oplus k\rangle}{\sqrt{2}}.
\end{eqnarray}

{\bf Decryption:} Suppose the key $k=k^{(1)}k^{(2)}\cdots k^{(l)}$, randomly choose one bit of $k$ whose value is 1. Here we assume the $j_{th}$ bit of $k$ is chosen. First we do controlled-NOT operation to each bit of ciphertext but the $j_{th}$ bit, the control bit of each operation is always the $j_{th}$ bit. Then we measure the $k_{th}$ bit with basis $|\pm\rangle$. If the result is $|+\rangle$ we have $x=0$, else $x=1$.

Note that here $k$ should not be a zero-string.

Similar to the proof in article\cite{Jia10,Hay08}, we denote the density operator of the ciphertext as $\rho_{(x,k)}$, then we have:
\begin{eqnarray}
\rho_{(x,k)}&=&\frac{1}{2\cdot 2^l}\sum_i(|i\rangle+(-1)^x|i\oplus k\rangle)(\langle i|+(-1)^x\langle i\oplus k|)\nonumber\\
&=&\frac{1}{2^l}\sum_i\sum_b(-1)^{bx}|i\rangle\langle i\oplus bk|
\end{eqnarray}
where $b\in\{0,1\}$.

While we encrypt a bit-string $x_1,x_2,\cdots, x_n$, the density operator of the ciphertext for attacker can be shown as
\begin{eqnarray}
\rho_{(x_1,x_2,\cdots,x_n)}=\frac{1}{2^l-1}\sum_k\rho_{(x_1,k)}\otimes\rho_{(x_2,k)}\otimes\cdots\otimes\rho_{(x_n,k)}
\end{eqnarray}

\begin{theorem}
If for every positive polynomial $p(\cdot)$, there exist a sufficiently large $N$ satisfy that while $n>N$, $\sqrt{\frac{1}{2^{l-n}}}<\frac{1}{p(n)}$, the protocol will be information-theoretically secure.
\end{theorem}

{\bf Proof:} Define $$\parallel X\parallel_{tr}=tr|X|.$$ For every $x=(x_1,x_2,\cdots, x_n)\in\{0,1\}^n$, $b=(b_1,b_2,\cdots, b_n)\in\{0,1\}^n$,$i=(i_1,i_2,\cdots, i_n)\in\{0,1\}^{l\times n}$,, we have
\begin{eqnarray}\label{rho1}
& &\parallel\rho_{(x_1,x_2,\cdots,x_n)}-(\frac{1}{2^{ln}}I_l^{\otimes n})\parallel_{tr}\nonumber\\
&=&\frac{1}{(2^l-1)\cdot2^{ln}}\parallel\sum_{b\neq\scriptsize\overrightarrow{0},k,i}(-1)^{b\cdot x}(|i_1,\cdots,i_n\rangle\langle i_1\oplus b_1k,\cdots,i_n\oplus b_nk|)\parallel_{tr}\nonumber
\end{eqnarray}

As $\parallel|w\rangle\langle v|\parallel_{tr}\leq\parallel|w\rangle\parallel_{tr}\parallel\langle v|\parallel_{tr}$ and $\sqrt{\frac{1}{2^{l-n}}}<\frac{1}{p(n)}$, we have:
\begin{eqnarray}\label{ineqn1}
& &\parallel\rho_{(x_1,x_2,\cdots,x_n)}-(\frac{1}{2^{ln}}I_l^{\otimes n})\parallel_{tr}\nonumber\\
&\leq&\frac{1}{(2^l-1)\cdot2^{ln}}\sum_i\Big(\parallel|i_1,\cdots,i_n\rangle\parallel_{tr}\cdot\parallel\sum_{b\neq\scriptsize\overrightarrow{0},k}(-1)^{b\cdot x}\langle i_1\oplus b_1k,\cdots,i_n\oplus b_nk|\parallel_{tr}\Big)\nonumber\\
&=&\frac{1}{(2^l-1)\cdot2^{ln}}\cdot2^{nl}\cdot\sqrt{(2^l-1)(2^n-1)}\leq\frac{1}{p(n)},
\end{eqnarray}

which means for every $x$,
\begin{eqnarray}
D(\rho_{(x)},\frac{1}{2^{ln}}I_l^{\otimes n})\leq\frac{1}{2}\times\frac{1}{p(n)}
\end{eqnarray}
so we can get the following inequality:
\begin{eqnarray}
D(\rho_{(x)},\rho_{(x')})\leq\frac{1}{p(n)}
\end{eqnarray}
Based on the Theorem.\ref{information}, we can say that the protocol is bound information-theoretically secure. $\Box$

We also assume the plaintext is the codewords being encoded with error correcting code(ECC) $C$ as above. Let the length of key is $m\times l$ and use every $l$-bit of the key as a group to encrypt one bit plaintext, it can used to encrypt $m$-bits plaintext once. When the length of the plaintext more than $m$ bits, we repeat this process and reuse the key.

The protocol B has a property as follow:

\begin{property}\label{per1}
While the attacker decrypts with a random bit-string except the unique key, the result will be 0 or 1 both with probability $\frac{1}{2}$.
\end{property}

{\bf Proof:}
Let $k=k^{(1)}\cdots k^{(s)}=1, k^{(s+1)}\cdots k^{(l)}=0$ is the unique key.

The ciphertext then can be expressed as
$$|i\rangle+(-1)^{b}|i\oplus k\rangle=|i^{(1)}\cdots i^{(n)}\rangle+(-1)^b|\bar{i}^{(1)}\cdots \bar{i}^{(s)}i^{(s+1)}\cdots i^{(l)}\rangle.$$

Let $k'=k'^{(1)}k'^{(2)}\cdots k'^{(l)}$ is another random bit-string. Without loss of generality, we assume $k'^{(1)}\cdots k'^{(t)}=1$, $k'^{(t+1)}\cdots k'^{(s)}=0$, $k'^{(s+1)}\cdots k'^{(r)}=1$, $k'^{(r+1)}\cdots k'^{(l)}=0$. Let the control bit chosen by him is the $j_{th}$ bit.

\begin{enumerate}
\item{\bf If $k_j=1$.}

For example, he chooses $k'^{(1)}$, and use the first bit of ciphertext as the control bit, there exists two cases.
\begin{enumerate}
\item
 If $i^{(1)}=0$, after the first step of decryption, the result is:
    \begin{eqnarray}
    & |i^{(1)}&\cdots i^{(l)}\rangle+(-1)^b|\bar{i}^{(1)}\cdots \bar{i}^{(s)}i^{(s+1)}\cdots i^{(l)}\rangle\stackrel{k'}{\longrightarrow}\nonumber\\
    & &|i^{(1)}\cdots i^{(l)}\rangle+(-1)^b|\bar{i}^{(1)}i^{(2)}\cdots i^{(t)}\bar{i}^{(t+1)}\cdots \bar{i}^{(s)}\cdots \bar{i}^{(r)}i^{(r+1)}\cdots i^{(l)}\rangle.\nonumber\\
  \end{eqnarray}

If and only if $t=s=r$, which means $k'=k$, he can make sure the value of $b$ when measuring the control bit with $|\pm\rangle$ basis, otherwise, there should exist another bit entangled with the control bit, then its reduced density operator can be expressed as
  \begin{eqnarray}
  \rho_1=\frac{1}{2}|0\rangle\langle0|+\frac{1}{2}|1\rangle\langle1|=\frac{1}{2}|+\rangle\langle+|+\frac{1}{2}|-\rangle\langle-|,
  \end{eqnarray}

so if he measures it with $|\pm\rangle$ basis, he will get 0 or 1 with same probability.

\item
If $i^{(1)}=1$, after the first step of decryption, the result is:
    \begin{eqnarray}
    & |i^{(1)}&\cdots i^{(l)}\rangle+(-1)^b|\bar{i}^{(1)}\cdots \bar{i}^{(s)}i^{(s+1)}\cdots i^{(l)}\rangle\stackrel{k'}{\longrightarrow}\nonumber\\
    & &|i^{(1)}\bar{i}^{(2)}\cdots\bar{i}^{(t)}i^{(t+1)}\cdots i^{(s)}\bar{i}^{(s+1)}\cdots \bar{i}^{(r)}i^{(r+1)}\cdots i^{(l)}\rangle+\nonumber\\
    & &+(-1)^b|\bar{i}^{(1)}\cdots \bar{i}^{(s)}i^{(s+1)}\cdots i^{(l)}\rangle.\nonumber\\
  \end{eqnarray}

The same as the first situation, the property follows.
\end{enumerate}

\item {\bf If $k_j=0$.}

For example, he chooses $k'^{(r+1)}$, and uses the $(r+1)_{th}$ bit of ciphertext as the control bit, there also exists two cases.

\begin{enumerate}
\item
If $i^{(s+1)}=0$, after the first step of decryption, the result is:
    \begin{eqnarray}
    &|i^{(1)}&\cdots i^{(l)}\rangle+(-1)^b|\bar{i}^{(1)}\cdots \bar{i}^{(s)}i^{(s+1)}\cdots i^{(l)}\rangle\stackrel{k'}{\longrightarrow}\nonumber\\
    & &|i^{(1)}\cdots i^{(l)}\rangle+(-1)^b|\bar{i}^{(1)}\cdots \bar{i}^{(s)}i^{(s+1)}\cdots i^{(l)}\rangle.
  \end{eqnarray}

While the control bit is $|0\rangle$ and independent from other bits here, the attacker measures it with $|\pm\rangle$ basis, he will get 0 or 1 with same probability.

\item
If $i^{(s+1)}=1$, the same as the above situation, after the first step of decryption, the control bit is $|1\rangle$ and independent from other bits.

Thus the property follows. $\Box$
\end{enumerate}
\end{enumerate}

This property means that once there exists any errors in one group, the decryption will result of 0 or 1 both with probability $50\%$, so we may just consider the number of groups with errors, named $m_0$. Let $p_e$ is the probability of verification of that if one $(m\times l)$-bit-string is the key.

Similarly as above, if $m_0<2t+1$, it turns out $n=-\frac{\ln(1-p_0)}{m_0\ln2}$; else it turns out $n=\frac{\ln(1-p_0)}{\ln(1-2^{2t-m_0})}$.

Then the unicity distance for $p_0$ is
\begin{eqnarray}
N_Q=n_0\times |K|,
\end{eqnarray}
here $|K|=2^{m\times l}$ is the size of the key space.

\subsection{Quantum encryption protocol with infinite unicity distance(QUD2)}
We know the quantum one time pad has infinite unicity distance based on its unconditional security, besides this we suggest in Sec. \ref{sec1} that the definition of QUD1 may lead to infinite unicity distance. Here we present another idea which also designs a quantum encryption protocol with infinite unicity distance under definition of QUD2. For this protocol, the ciphertext with any length can not be used to check more than one bit-string. The protocol is shown as follow:

The plaintext $X$ is a classical bit-string with length $n(n=1,2,3\ldots)$, $X=x_1,\cdots,x_n$, $x_i\in{0,1}$.
\begin{enumerate}
  \item
  Firstly $X$ is encoded into $|\phi_X\rangle=|x_1\rangle\cdots|x_n\rangle$. Its vector notation is
  \begin{eqnarray}
  |\phi_X\rangle\equiv\left[\begin{array}{cc}\bar{x}_1\\x_1\end{array}\right]\otimes\cdots\otimes\left[\begin{array}{cc}\bar{x}_n\\x_n\end{array}\right]
  =\left[\begin{array}{cc}\bar{x}_1\bar{x}_2\cdots\bar{x}_n\\\vdots\\x_1x_2\cdots x_n\end{array}\right].
  \end{eqnarray}
  only one row of the last vector is 1 and the other $2^n-1$ rows is 0.

  \item
  Then the protocol may generate a quantum operator $\mathcal{E}_{k,n}$ determined by the key $k$ and the length $n$:
  \begin{eqnarray}
  \mathcal{E}_{k,n}\equiv\left[\begin{array}{ccc}
                                 a_{1,1}(k) & \cdots & a_{1,2^n}(k)  \\
                                 \vdots & \ddots & \vdots \\
                                 a_{2^n,1}(k) & \cdots & a_{2^n,2^n}(k)
                               \end{array}\right].
  \end{eqnarray}

  \item The ciphertext $|\phi_Y\rangle=\mathcal{E}_{k,n}(|\phi_X\rangle)$ is a qubit-string. Assume the $l_{th}$ row of $|\phi_X\rangle$ is 1, then the vector notation of $|\phi_Y\rangle$ can be shown as:
  \begin{eqnarray}
  |\phi_Y\rangle\equiv\left[\begin{array}{c}a_{1,l}(k)\\\vdots\\a_{2^n,l}(k)\end{array}\right].
  \end{eqnarray}
\end{enumerate}

They satisfy that each qubit of $|\phi_Y\rangle$ should be entangled with others.

According to this kind of protocol, the length of $k$ is not increase with increasing of plaintext, and the $n$ is said to be open for the attacker, so the number of keys that need exhaustive is remain the same. It means this protocol is different from one-time pad protocol. On the other hand,  Bob has the right key $k$ and is able to decrypt the ciphertext. Meanwhile, once the attacker Eve tries to decrypt the ciphertext with any bit-string $k'$, all the qubits will collapse, which means no matter how long the plaintext is, Eve can only check that if one bit-string $k'$ is the key. In other words, this kind of protocol also has infinite unicity distance.

The key point to this kind of protocol is that, how to design a scheme as follow: the scheme can generate a quantum operator $\mathcal{E}_{k,n}$ that can be applied to a qubit-string with length $n$ for every $n$, $\mathcal{E}_{k,n}$ should be determined by $k$ with limited length, and each column of the matrix of $\mathcal{E}_{k,n}$ cannot do direct product decomposition.

In fact, the last condition can be changed into that the matrix of $\mathcal{E}_{k,n}$ cannot do direct product decomposition.

On the other hand, there exists some properties that may reduce the unicity distance for quantum encryption protocols. If a quantum encryption protocol for whom every ciphertext decrypted with any key will result a computational basis, its unicity distance can be reduced greatly. Because we can use one ciphertext string to check every key again and again with this kind of quantum encryption. The existence of this encryption has not been proved, but all the classical encryptions belongs to this group in a sense.

\subsection{Discussions}
Our analysis of unicity distance of CCQ encryption protocol above is only for one kind of quantum encryption protocols, the unicity distance of other three kinds of quantum encryption protocol will be discussed in the further research. We derived the unicity distance according to the given quantum encryption protocols, these results need some precondition such as the expression of plaintext space, so how to derive the unicity distance of a formalized quantum encryption protocol is still an open problem.

Note that there is a big gap between the condition of the Private Quantum Channel and the unicity distance of the given protocols, this would be a normal behavior, so how can we define a protocol is secure in practice(not perfect secure) is worth to discuss.

\section{Conclusion}
We classify encryption protocols into five types, then we present a definition of the unicity distance of CCQ encryption protocol(the encryption protocols whose plaintext and key are classical but algorithm is a quantum one). We show that we can not completely copy the proof of classical unicity distance to prove quantum one, then we present three useful definitions of quantum unicity distance.

In classical context, a encryption protocol must have unicity distance while the key is reused and the plaintext has nonzero redundancy, so we can tell that only one time pad can lead to infinite unicity distance while a language always has nonzero redundancy. In Quantum context, quantum one time pad(PQC) also has infinite unicity distance, we suggest that based on QUD2 there may exist other quantum encryption protocols which also has infinite unicity distance besides quantum one time pad, it is very different from classical one.

\section*{Acknowledgment}
This work was supported by the National Natural Science Foundation of China (Grant No. 61173157).




\bibliography{reportt}   
\bibliographystyle{spiebib}   







\end{document}